\documentstyle[aps,epsfig,float,twocolumn]{revtex}

\begin{document}

\title{Bilayer Quantum Hall Systems at Filling Factor $\nu=2$:\\
An Exact Diagonalisation Study}
\author{John Schliemann$^{\ast}$\cite{emadd}
and A.~H. MacDonald$^{\dag}$}
\address{$^{\ast}$Physikalisches Institut, Universit\"at Bayreuth,
D-95440 Bayreuth, Germany\\
$^{\dag}$Department of Physics, Indiana University, 
Bloomington, IN 47405}
\date{\today}
\maketitle
\begin{abstract}
We present an exact diagonalisation study of bilayer quantum Hall systems
at filling factor $\nu=2$ in the spherical geometry. We find the
high-Zeeman-coupling phase boundary of the broken symmetry canted
antiferromagnet is given exactly by previous Hartree-Fock mean-field
theories, but that the state's stability at weak Zeeman coupling
has been qualitatively overestimated.  In the absence of
interlayer tunneling, degeneracies occur between total spin
multiplets due to the Hamiltonian's invariance under independent
spin-rotations in top and bottom two-dimensional electron layers.
\end{abstract}

\vspace{0.5cm}

In the last decade there has been an increasing interest in
quantum Hall ferromagnets \cite{DaPi:97}. Most recently
bilayer systems at a filling factor of two have become the object of
intensive theoretical \cite{ZRS:97}--\cite{YaCh:99} and experimental 
\cite{PPDPPW:97}--\cite{KDSDHWCG:99}
research. 
The rich phenomenology of $\nu=2$ bilayers mirrors a complex 
interplay between Coulomb interactions in the lowest Landau level,
Fermi statistics, and the coupling of external fields to 
spin and layer degrees of freedom.\\
Our current microscopic understanding of
$\nu=2$ bilayer quantum Hall ferromagnets is based on Hartree--Fock
mean--field theory calculations \cite{ZRS:97,MRJ:99}, and on
a partially phenomenological effective spin Hamiltonian
description \cite{DeDa:99}. Both approaches lead to the prediction of a novel
broken symmetry canted--antiferromagnet ground state with finite 
spin--suscpetibility which interpolates, as external field 
parameters are varied,
between a fully spin-polarized state and a spin singlet state,
both of which have charge and spin gaps and zero differential 
spin susceptibility.
In this Letter we report on the first exact diagonalisation study
of finite bilayer systems at filling factor $\nu=2$.  Our
calculations support the predicted occurrence of a broken symmetry
ground state, and indeed demonstrate that the Hartree--Fock result for
its large Zeeman coupling phase boundary is exact. We find that the
stability of the canted antiferromagnet state relative to the
spin singlet state is overstated by the Hartree--Fock
approximation, and estimate the correct position of the phase
boundary.\\
A bilayer system  in a strong magnetic field is described in spherical 
geometry by the following Hamiltonian,
\begin{equation}
{\cal H}={\cal H}_{{1\rm P}}+{\cal H}_{{\rm Coul}}\quad,
\end{equation}
where ${\cal H}_{{\rm Coul}}$ represents the usual Coulomb interaction
within and between layers, and the single--particle Hamiltonian 
${\cal H}_{{1\rm P}}$ is given by
\begin{eqnarray}
{\cal H}_{{1\rm P}} & = &
-\frac{1}{2}\sum_{m}c^{+}_{\mu,\sigma,m}
\big[\Delta_{v}\tau^{z}_{\mu,\mu'} \delta_{\sigma,\sigma'}\nonumber\\
& &+\Delta_{t}\tau^{x}_{\mu,\mu'}\delta_{\sigma,\sigma'}
+\Delta_{z}\delta_{\mu,\mu'}\sigma^{z}_{\sigma,\sigma'}\big]
c_{\mu',\sigma',m}\,.
\end{eqnarray}
A summation convention is understood for repeated greek indices, where 
$\mu,\mu'\in\{+,-\}$ run over the layer (or pseudospin) index, while
$\sigma,\sigma'\in\{\uparrow,\downarrow\}$ run over the $z$--projections of
the electron spin; $\vec\tau$, $\vec\sigma$ are Pauli matrices for
pseudospin and electron spin, respectively. 
$m\in\{-N_{\phi}/2,\dots,N_{\phi}/2\}$
is the z-projection of the orbital angular momentum of each electron
in the lowest Landau level, 
where $N_{\phi}$ is the number of flux quanta penetrating the sphere.
The Hamiltonian contains bias voltage
($\Delta_{v}$), tunneling ($\Delta_{t}$), and
Zeeman coupling ($\Delta_{z}$) terms.  In the following we
measure the interlayer separation $d$ in units of the magnetic length
$l_{B}=\sqrt{{\hbar c}/{eB}}$ and all energies in units of the
Coulomb energy scale $e^{2}/\epsilon l_{B}$.\\
We first consider the case where all single--particle coupling constants 
vanish.
The number of particles in each layer is then a good quantum number and, in the
ground state, both layers have filling factor one.  Moreover,
because the Coulomb
interaction is spin--independent, the Hamiltonian is invariant under
independent spin rotations in either layer. 
It follows that the total spin in
either layer is a good quantum number.

\begin{figure}
\begin{center}
\epsfig{file=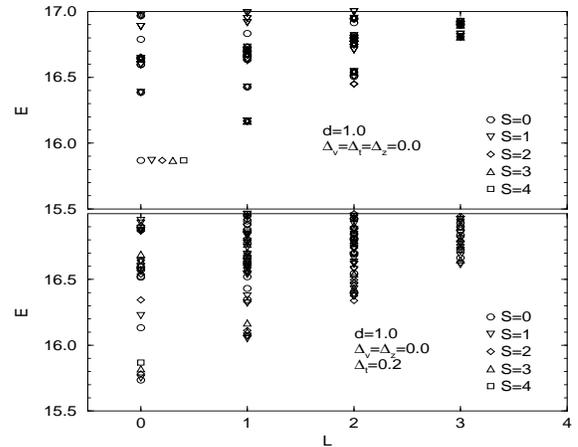,height=0.25\textheight,width=0.99\linewidth}
\caption{The low--lying spectrum of a double layer system at filling factor
two with eight electrons. The energy levels are plotted as a function of
the total orbital angular momentum $L$ for different values of the total
spin $S$. The upper diagram shows the case of vanishing
single--particle couplings, while in the lower spectrum $\Delta_{t}$ is finite.
For $\Delta_{t}=0$, the degenerate spin multiplets in the
ground state are exhibited explicitly .
\label{fig1}}
\end{center}
\end{figure}
For layer separation $d \to \infty$
it is known that the ground state of each isolated layer is a quantum Hall
ferromagnet with $S=N/4$, where $N$ is the total number of particles in the 
two layers. Since the Coulomb interaction 
within the layers is stronger than between them,
we anticipate that this should remain true at any finite $d$.
Indeed, this expectation is confirmed numerically.
The upper diagram of figure \ref{fig1} shows the low--lying spectrum of a
double layer system at filling factor two with eight electrons in the case
of vanishing single--particle couplings. The ground state as
described there consists of multiplets with total spin quantum number
$S$ varying from $0$ to $4$, the total spin representations of a
state with good spin quantum number $S=2$ in each layer.  When
an interlayer tunneling term is added to the Hamiltonian, the
additional symmetry responsible for these degeneracies is lost and
only total spin is a good quantum number.\\
The robustness of individual layer ground state spin quantum
numbers against the effects of added interlayer interactions
is not restricted to total filling factor two.
In fact, in our finite-size numerical calculations,
it holds at all combination of numbers of flux quanta and electrons we have 
checked.

\begin{figure}
\begin{center}
\epsfig{file=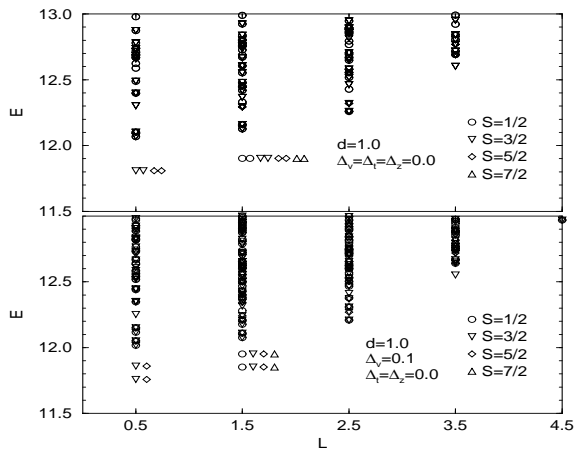,height=0.25\textheight,width=0.99\linewidth}
\caption{The same system as in figure \protect\ref{fig1}, but with one
electron removed. The degenerate multiplets discussed in the text are
shown in detail.
\label{fig2}}
\end{center}
\end{figure}

For instance, consider the system discussed above, but with one electron
either added or removed.  (These two cases are equivalent by
particle--hole symmetry.) In the ground state, one of the
layers has a filling factor of one (a ground state with $L=0$,
$S=2$), while the other one contains a hole and the low--lying states
are organised in the well--known skyrmion branch having quantum numbers
$L=S=1/2,3/2$ \cite{XiHe:96}.
The spectrum of this double layer system is shown in
the upper diagram of figure \ref{fig2}. The ground state carries quantum
numbers $L=1/2$ and $S=3/2$, $S=5/2$ and is the result of coupling the
$L=0$, $S=2$ ground state of the full layer to the $L=S=1/2$ ground state
of the hole system. Since either layer can carry the hole,
we get two copies of the degenerate multiplets. The next higher degenerate
group of multiplets is the result of coupling the ground state of the $\nu=1$
layer to the $L=S=3/2$ state of the hole layer, and again all multiplets
are doubled. To confirm this interpretation we have verified that that this degeneracy
doubling is lifted by applying a bias voltage, as shown in the lower diagram of
figure \ref{fig2}.   
These results demonstrate that in the absence
of interlayer tunneling, bilayer states for $\nu$ near $2$ can be safely
regarded as two single-layer quantum Hall ferromagnets whose coupling
has only a quantitative significance, for example in changing the energies of the
skyrmionic elementary charged excitations. 

With this established, we now focus on changes in the nature of the 
ground state at $\nu=2$ as the single--particle coupling constants 
are varied.    
An arbitrarily small tunneling amplitude is enough to break the degeneracy 
among the different spin multiplets and make the spin singlet state with the 
most antiparallel electron spin structure and consequently the most parallel 
pseudospin structure the nondegenerate ground state.
More precisely, the energy levels are found to be ordered by the total spin
$S$ with the difference between neighboring levels increasing with increasing
$S$, as shown in the lower diagram of figure \ref{fig1}.
We note that the lone maximally polarized $S=N/2$ multiplet is 
annihilated by the interlayer tunneling term in the Hamiltonian and has an
eigenenergy which is independent of $\Delta_t$.
Turning on the Zeeman coupling $\Delta_{z}$ at a given value
of the tunneling $\Delta_{t}$ does not change the eigenstates themselves, but
only shifts their energies and breaks the degeneracy {\em within} each spin
multiplet. 

These findings lead to the following scenario: With increasing
$\Delta_{z}$ a lower critical Zeeman coupling 
$\Delta_{z}^{(1)}(\Delta_{t})$ 
is reached where the state with $S^{z}=S=1$ becomes the ground 
state, i.~e. the system leaves the spin singlet phase. If $\Delta_{z}$ is 
increased further, the ground state $S^{z}=S$ quantum number increases  
monotonously until, at an upper critical value, 
$\Delta_{z}^{(2)}(\Delta_{t})$ 
the fully spin--polarized state $S^{z}=S=N/2$ is reached. 

\begin{figure}
\begin{center}
\epsfig{file=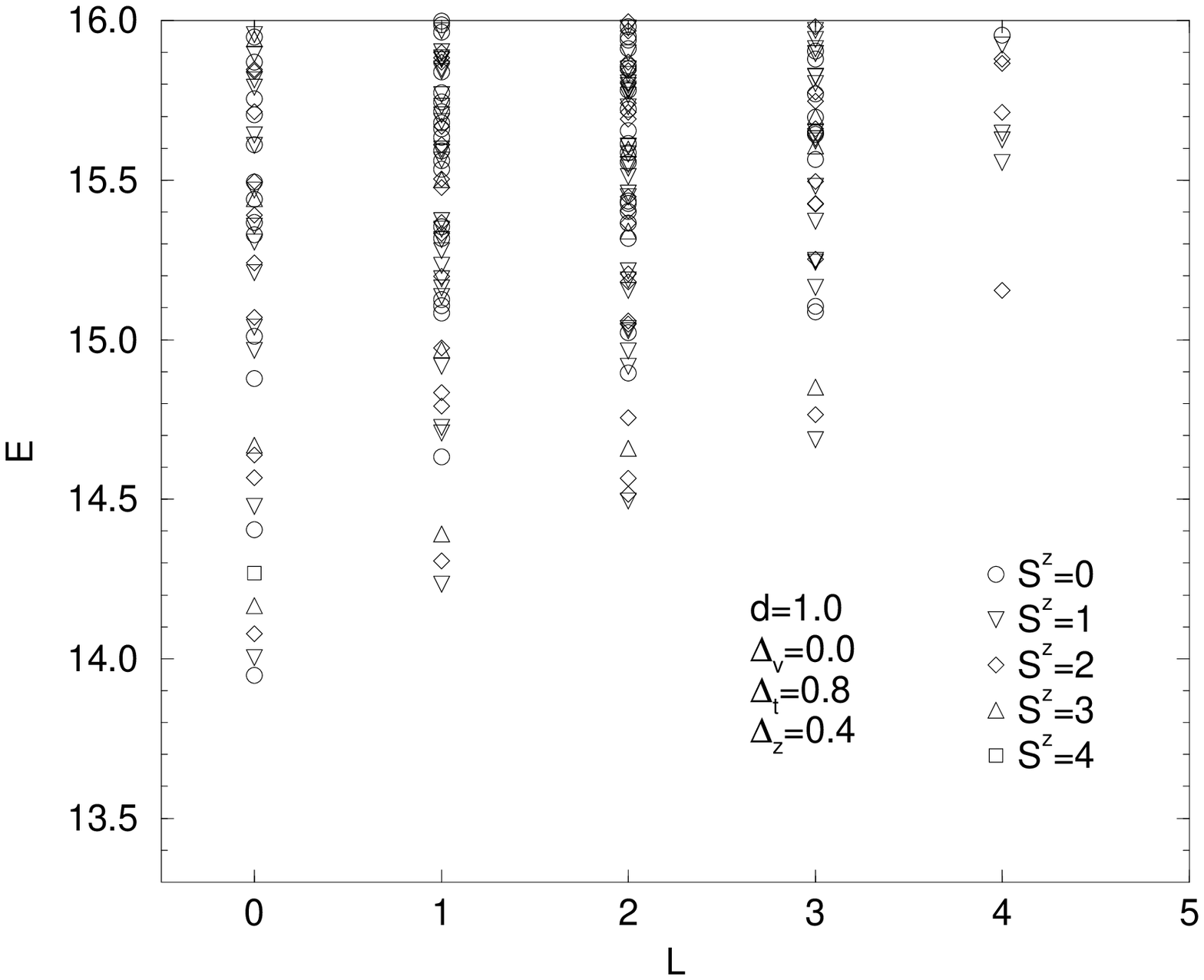,height=0.25\textheight,width=0.99\linewidth}
\epsfig{file=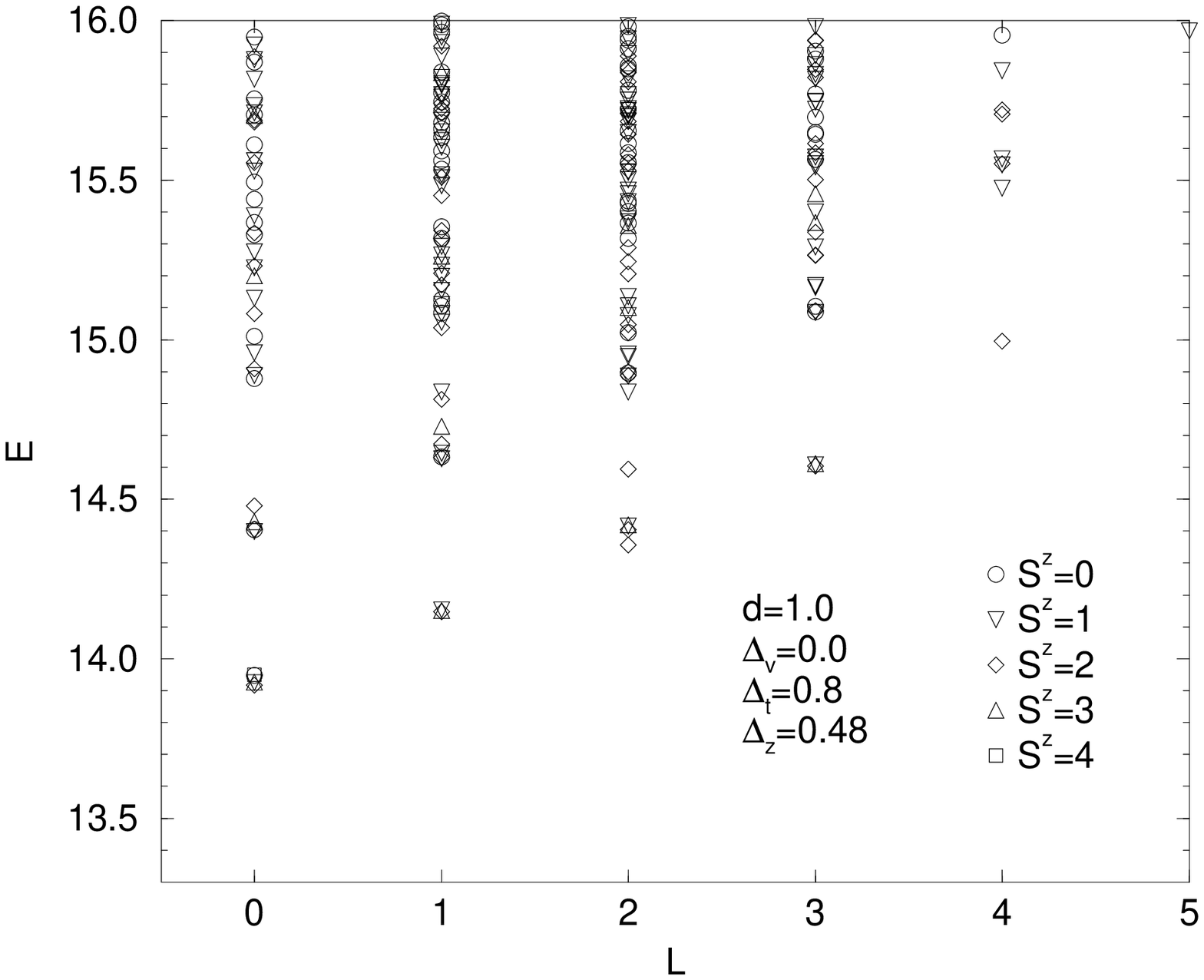,height=0.25\textheight,width=0.99\linewidth}
\epsfig{file=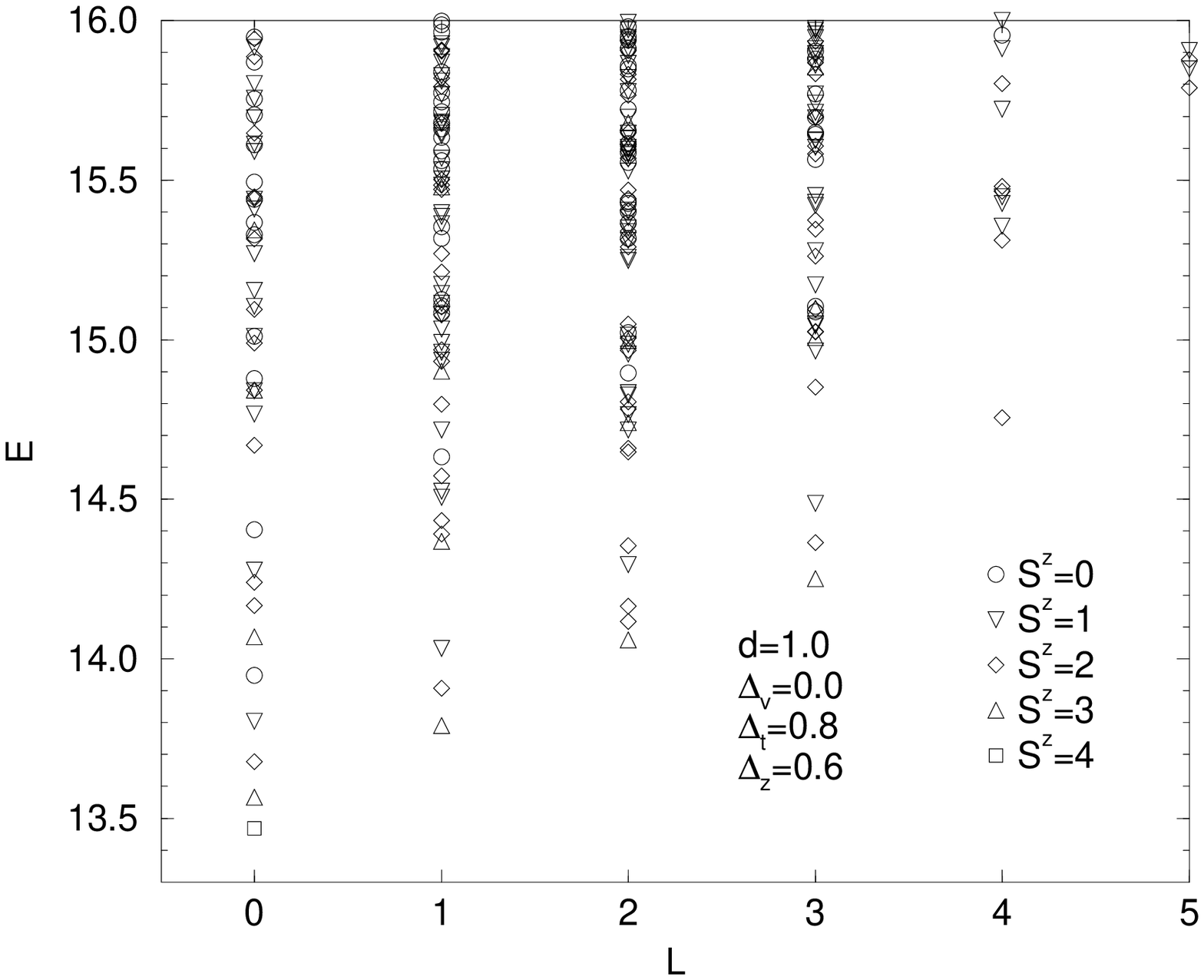,height=0.25\textheight,width=0.99\linewidth}
\end{center}
\caption{Finite-size spectra for of a system of eight electrons
as a function of the Zeeman coupling for 
tunneling amplitude $\Delta_{t}=0.8$.  States with negative values of 
$S^{z}$ are omitted for clarity. 
\label{fig3}}
\end{figure}

Finite--size spectra are shown in figure \ref{fig3}
for several Zeeman couplings at $\Delta_{t}=0.8$. In the
top diagram the system is in the spin singlet phase, while in the 
bottom the ground state is fully spin--polarized. In the narrow transition
area $\Delta_{z}^{(1)}\leq\Delta_{z}\leq\Delta_{z}^{(2)}$ all low--lying
states with $S^{z}=0,\dots,N/2$ have energies very close to each other.
This holds also for a branch of states with angular momentum
$L>0$ which appears to be separated by a gap from higher-lying parts of the  
spectrum. We identify this transition region with the 
canted antiferromagnetic phase first proposed by Zheng {\it et al.} 
on the basis of the unrestricted Hartree--Fock approximation\cite{ZRS:97}.

The mean-field state, which breaks 
spin--rotational symmetry around the $\hat z$ axis can be constructed 
as a linear combination of the nearly degenerate exact
eigenstates $|S^{z}\rangle$ carrying quantum numbers $L=0$ and definite 
$S^{z}$ values. To analyse the spin structure perpendicular
to the Zeeman axis we introduce spin operators for each layer
separately, $\vec S_{\mu}$,
$\mu\in\{+,-\}$, where the total spin of the bilayer system is given by
$\vec S=\sum_{\mu}\vec S_{\mu}$. 
Since the states $|S^{z}\rangle$ belong to different multiplets, all matrix 
elements of $S^{\pm}$ between them are zero, which means that 
\begin{equation}
\langle S^{z}|S_{\mu}^{+}|S^{z}-1\rangle
=-\langle S^{z}|S_{-\mu}^{+}|S^{z}-1\rangle\,.
\end{equation}
Thus, the matrix elements of spin components perpendicular to the
Zeeman axis have the same magnitude and opposite sign in opposite layers. 
Any wavepacket constructed from these states will, like the mean-field state,
have opposite transverse spin--polarization in the two layers. \\

\begin{figure}
\begin{center}
\epsfig{file=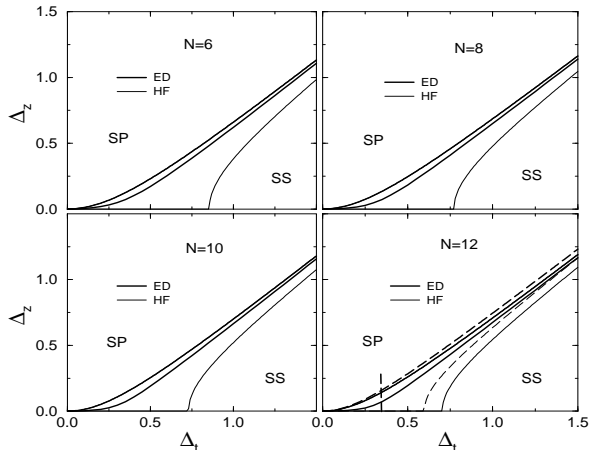,height=0.25\textheight,width=0.90\linewidth}
\caption{Phase diagrams at a layer separation of $d=1.0$ and zero bias voltage
for different system sizes. The upper and lower critical Zeeman coupling
as obtained from exact diagonalisation (ED) data are plotted as thick lines.
The canted antiferromagnetic phase lies between them and separates the
spin--polarized phase (SP) from the spin singlet phase (SS). The finite-size
Hartree-Fock (HF) results are given for comparision. The upper HF phase
boundary coincides with the ED result (see text), while the lower phase
boundary strongly overestimates the region of the canted antiferromagnetic
ground state.\newline
The last panel ($N=12$) shows in addition the the HF phase boundaries for
the infinite system (where the upper one is as well exact) as dashed lines.
The value of $\Delta_{t}$ below which the ED results for 
$\Delta_{z}^{(1)}(\Delta_{t})$ decrease with system size is marked by a 
vertical line.
\label{fig4}}
\end{center}
\end{figure}

Next let us consider the phase diagram, i.~e. the functions
$\Delta_{z}^{(1)}(\Delta_{t})$ and $\Delta_{z}^{(2)}(\Delta_{t})$. 
In the thermodynamic limit these lines mark the boundaries between 
spin--polarized, spin--singlet, and canted antiferromagnet phases.
They may be compared with the Hartree--Fock results
obtained recently in the planar geometry for an infinite system 
\cite{MRJ:99}. To avoid unnecessary finite--size uncertainty, 
we have rederived the Hartree--Fock equations for the
spherical geometry obtaining explicit expressions for finite systems.
We find that the phase boundaries have the same form as for the infinite
system \cite{MRJ:99}:
\begin{equation}
\Delta_{z}^{(1)}=\sqrt{\Delta_{t}(\Delta_{t}-2F_{-})}
\label{pb1}
\end{equation}
for $\Delta_{t}>2F_{-}$, otherwise $\Delta_{z}^{(1)}=0$, and
\begin{equation}
\Delta_{z}^{(2)}=\sqrt{\Delta_{t}^{2}+F_{-}^{2}}-F_{-}\quad.
\label{pb2}
\end{equation}
In the present case, however, the exchange parameter $F_{-}$ is 
size--dependent:
\begin{equation}
F_{-}=\frac{e^2}{\epsilon l_{B}}\frac{N_{\phi}+1}{\sqrt{2N_{\phi}}}
\left(I(1)
-\left(\frac{1}{\alpha}\right)^{N_{\phi}+\frac{1}{2}}I(\alpha)\right)\quad,
\label{Fminus}
\end{equation}
with
\begin{equation}
I(\alpha)=\int_{0}^{\alpha}\,dx\frac{x^{N_{\phi}}}{\sqrt{1-x}}
\quad,\quad
\alpha=\frac{1}{1+\frac{1}{N_{\phi}}\frac{d^2}{2l_{B}^{2}}}\,.
\label{intpar}
\end{equation}
In figure \ref{fig4} we compare exact--diagonalization
and Hartree--Fock finite-size phase boundaries 
for $N=6,8,10,12$ electrons.  Not unexpectedly, the
region of the canted antiferromagnetic phase turns out to be much smaller
than predicted by the Hartree--Fock theory.
Interestingly however, the two results for the phase boundary $\Delta_{z}^{(2)}(\Delta_{t})$
coincide within our numerical precision of at least $10^{-12}$.
Since both the spin--polarized
state and its elementary collective excitations, 
which go soft at the phase boundary, are 
described exactly \cite{Mac:85} by Hartree--Fock theory, this coincidence is 
not entirely unexpected.  It does, however, very convincingly demonstrate that
the spin-polarized to canted phase transition remains continuous when quantum 
fluctuations are 
included.  We conclude that the position of this phase boundary 
in the thermodynamic limit can be calculated {\em exactly} using 
the expressions given in reference 
\cite{MRJ:99}, or equivalently the $N_{\phi}\to\infty$ limit of 
equations (\ref{pb2})--(\ref{intpar}), adding finite--well--width and 
other sample--specific corrections as required.

The result for this upper boundary is in marked contrast with our findings
for the lower boundary. Here the Hartree--Fock approach leads 
to a canted antiferromagnetic region at all system sizes 
even in the absence of the Zeeman coupling.
A similar conclusion was reached by  
Demler and Das Sarma using the effective spin theory \cite{DeDa:99}.
In our finite--size exact diagonalization calculations, on the other hand,
$\Delta_{z}^{(1)}$ is always finite.  For $\Delta_{t} \lesssim 0.35$, marked
by a vertical line in the last panel of figure ~\ref{fig4}, $\Delta_{z}^{(1)}$ 
decreases monotonously with system size, while it increases monotonously
for larger values of $\Delta_{t}$. These observations guarantee that
the canted antiferromagnetic phase predicted by Hartree--Fock theory is
actually present in the infinite system (although clearly diminished
by quantum fluctuations). Moreover, our findings are  
consistent with a vanishing thermodynamic limit of 
$\Delta_{z}^{(1)}(\Delta_{t})$ and a nonzero spin susceptibility
for $\Delta_{t} \lesssim 0.35$. Static correlation function calculations
\cite{unpub}
at $\Delta_{z}=0$ are also consistent with a nonzero canted order parameter
for $\Delta_{t} < 0.35$.  These numerical results are thus consistent with 
a single intermediate phase which has both a finite spin--susceptibility and 
canted--antiferromagnet order.  For $\Delta_{t} \gtrsim 0.35$, 
$\Delta_{z}^{(1)}$ increases slowly with system size, presumably saturating at 
a finite value smaller than $\Delta_{z}^{(2)}$ and leaving a narrow 
canted-antiferromagnet strip in the phase diagram.
The maximum $\Delta_{t}$ value at which the ordered state phase extends down
to $\Delta_{z}=0$ is much smaller than the value $\Delta_{t}=0.60$, predicted
by Hartree--Fock theory.  
These findings are illustrated along with the exact upper phase boundary
in the right bottom panel of figure \ref{fig4}, which shows that
the Hartree--Fock approximation strongly overestimates the stability of 
the canted antiferromagnetic phase against the spin singlet phase
in finite systems and as well in the thermodynamic limit.\\
JS acknowledges support from the Deutsche Forschungsgemeinschaft under
grant SCHL 539/1--1.
AHM was supported from the National Science Foundation under 
grant DMR--9714055.

\end{document}